\documentclass[prl,twocolumn,english,showpacs,superscriptaddress]{revtex4}
\usepackage[latin9]{inputenc}
\usepackage{amsmath,amssymb,amsfonts}
\usepackage{CJK}
\usepackage{bm}
\usepackage{tikz}
\usepackage{xcolor}
\usetikzlibrary{matrix,fit}
\usepackage{textcomp}
\DeclareGraphicsExtensions{.pdf,.png,.jpg}

\usepackage{color}

\usepackage{babel}
\usepackage{slashed}

\newcommand{\avg}[1]{\left< #1 \right>}
\newcommand{\bra}[1]{\big<#1|}
\newcommand{\ket}[1]{|#1\big>}
\newcommand{\ZZ}{\mathbb{Z}}
\newcommand{\ra}{\rightarrow}

\newcommand{\ua}{\uparrow}
\newcommand{\da}{\downarrow}
\newcommand{\comment}[1]{}

\begin{document}

\title{Majorana modes meet fractional fermions in one dimension}

\author{Dan-bo Zhang}
\email[]{dbzhang@hku.hk}
\affiliation{Department of Physics and Center of Theoretical and Computational Physics, The University of Hong Kong, Pokfulam Road, Hong
Kong, China}

\author{Qiang-Hua Wang}
\email[]{qhwang@nju.edu.cn}
\affiliation{National Laboratory of Solid State Microstructures $\&$ School of Physics, Nanjing University, Nanjing, 210093, China}
\affiliation{Collaborative Innovation Center of Advanced Microstructures, Nanjing University, Nanjing 210093, China}

\author{Z. D. Wang}
\email[]{zwang@hku.hk}

\affiliation{Department of Physics and Center of Theoretical and Computational Physics, The University of Hong Kong, Pokfulam Road, Hong
Kong, China}

\date{\today}
\begin{abstract}
Majorana modes and fractional fermions are two types of edge zero modes appearing separately in topological superconductors and dimerized
chains. Here we reveal how to harvest both types of edge modes simultaneously in an exotic chain. Such modes are naturally spin-charge
separated, and are protected by the inversion and spin-parity symmetries. We construct a lattice model to illustrate the nature of these edge
modes, utilizing fermionic functional renormalization group, mean-field theory and bosonization. We also elucidate that the four-fold
degenerate ground states with edge-spinons in the Haldane phase of spin-$1$ chain may be reinterpreted as our spin-charge separated edge
modes in an equivalent spin-$1/2$ fermionic model.
\end{abstract}

\maketitle

\textit{Introduction}
Topological gapped fermionic systems are characterized by a finite energy gap in the bulk and topologically protected gapless states on the
boundary~\cite{WXG,classfication,ZYX-edge}. In one dimensional (1D) topological systems, the edge zero modes may be understood as a
consequence of the so-called symmetry-protected topological (SPT) fractionalization~\cite{Chen}. For example, the presence of Majorana zero
modes (MZMs) in a 1D topological superconductor (TSC)~\cite{Kitaev} is a result of SPT fractionalization of $\ZZ_2$ fermion parity, while the
emergence of fractional fermion (FF) (with a fractional charge) in a dimerized chain, the analogue of Su-Schrieffer-Heeger (SSH) model for
polyacetylene~\cite{SSH}, stems from the SPT fractionalization of inversion symmetry $P_I$. On the other hand, a unique property of 1D
fermionic spinful system is  spin-charge separation~\cite{Giamarchi}, or decoupling of charge and spin degrees of freedom. Thus an intriguing
and fundamental question arises:  whether topological bulk or/and edge states can be realized in charge and spin channels independently in
the same 1D SPT system?

To answer the question, we need to consider spinful systems and how edge zero modes emerge therein. One remarkable example is a DIII class
topological superconductor (DSC) with spin rotational $U(1)$ symmetry~\cite{Zhao-MF}. It is the fractionalization of the $\ZZ_2$ parity
symmetry (for charges) that results in MZMs at the edges.  Due to the time-reversal symmetry, MZMs at each end form time-reversal partners
and  carry opposite spins. This inspires us to realize SPT fractionalization in the spin and charge channels independently, such that
edge zero modes would appear independently and therefore inherit spin-charge separation in the bulk. Moreover, the edge modes in the spin
channel could be restructured as Majorana modes if a parity symmetry in the spin channel could be implemented. In this way, it appears
possible to realize an exotic type of edge zero modes composed of Majorana modes in the spin channel and fractional fermions in the charge
channel.

In this paper, we demonstrate how to realize the above-mentioned exotic edge zero modes. We propose a 1D lattice model of interacting
fermions with inversion symmetry and spin parity symmetry. Using functional renormalization group as a guide, we reach a mean field theory
with a bond-centered spin-density-wave (bSDW) order. The ground state is 4-fold degenerate, leading to four edge zero modes. By
fractionalization of the spin parity, these modes may form a product of Majorana modes in the spin channel and fractional fermion
modes in the charge channel. The results are corroborated by an effective field theory based on bosonization. The unique structure of edge
zero modes uncovered here enables separate manipulation of spin and charge degrees of freedom, which may be desirable in topological
qubits~\cite{Alicea}. Finally, we present an alternative theoretical understanding of the Haldane phase of spin-1 chain based on the above
results.\\

\textit{Lattice model}
Let us begin with a one dimensional lattice model of spinful fermions,
\begin{eqnarray}
H_{lat}=\sum_{j\sigma}[-t(c_{j\sigma}^\dagger c_{j+1\sigma}+{\rm h.c})-\mu c_{j\sigma}^\dagger c_{j\sigma}]   \nonumber\\
+ W_1 \sum_{j}(c_{j\uparrow}^\dagger c_{j\downarrow}^\dagger c_{j+1\downarrow}c_{j+1\uparrow}+{\rm h.c}) \nonumber\\
+  W_2\sum_{j}(c_{j\ua}^\dagger c_{j+1\ua}^\dagger c_{j+1\da} c_{j\da}+{\rm h.c}),
\label{model-lattice}
\end{eqnarray}
where $c_{j\sigma}$ annihilates an electron of spin $\sigma$ at site $j$, and $W_1$ ($W_2$) is the interaction strength for site-wise
singlet-pair hopping~\cite{pair-hopping-W1} (bond-wise triplet-pair spin flipping). Throughout this paper we focus on half filling so that we
set $\mu=0$. The system respects translation symmetry, inversion symmetry, particle-hole symmetry and time-reversal symmetry.
The triplet-pair spin-flipping breaks the global spin SU(2) symmetry, but the total spin component $S_z$ changes only by multiples of $\pm
2$, leaving a discrete spin symmetry characterized by a $\ZZ_2$ spin parity $P_S=(-1)^{S_z}$ for even-number sites. The inversion symmetry is known to protect
fractional fermions in dimerized chain, and the spin parity is a key ingredient in realizing topological superconductivity in the absence of
superconducting reservoir~\cite{Cheng-meng,Kraus}.\\

\textit{FRG-guided mean field theory}
We limit ourselves to repulsive interactions $W_1>0$ and $W_2>0$ throughout this paper. Both interactions would promote the bSDW
order~\cite{BSDW}, $\langle c_j^\dag \sigma_x c_{j+1}+{\rm h.c.}\rangle \sim (-1)^j M $. (Henceforth $\sigma_{x,y,z}$ are Pauli matrices in
the spin basis, and $c_j$ is a two-component spinor.) Such a state breaks the A-B sublattice symmetry and mixes spin parities, forming the
suitable basis for SPT fractionalizations. However, there are other competing orders. For example, a repulsive $W_1$ would
promote singlet pair-density-wave (sPDW), $\langle c_{j\ua}^\dag c_{j\da}^\dag \rangle \sim (-1)^j\Gamma$, while a repulsive $W_2$ would
promote site-local SDW, $\langle c_j^\dag \sigma_x c_j \rangle \sim (-1)^j m$, as well as a particular triplet pairing, $\langle c_j^\dag
\sigma_x i\sigma_y (c_{j+1}^\dag)^t\rangle \sim \Delta$. To treat all potential orders on equal footing, we resort to the singular-mode FRG
\cite{smfrg}. It turns out that for $W_1/W_2>1/8$, the bSDW is the only instability of the normal state at low energy scales. (More details
can be found in the Appendix~\ref{sm-frg}) Therefore, in this parameter space the low-energy physics can be safely described by an effective mean field
hamiltonian,
\begin{equation} H_{MF} = \sum_{j} \left\{ c_j^\dag [-t\sigma_0 + (-1)^j M \sigma_x] c_{j+1} + {\rm h.c.} \right\},\label{mft}
\end{equation}
where $\sigma_0$ is the identity matrix in spin basis. The mean field hamiltonian has an emerging symmetry that conserves the spin component
$S_x$. Along this quantization axis, $\sigma_x$ is diagonal (with eigenvalues $\pm 1$), and $H_{MF}$ is manifestly a doubled version of the
SSH model.\\

\begin{figure}
\centering
\includegraphics[width=8cm,height=2.5cm]{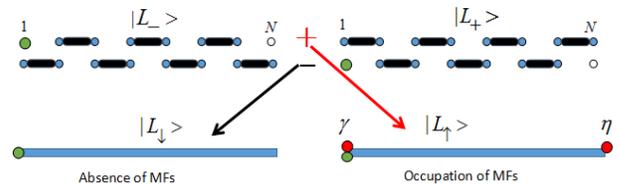}

\caption{Edge zero modes of fractional fermions (green dot) and Majorana fermions ($\gamma_L$ and $\gamma_R$, red dots). The system is
composed of two rows of dimerized chains (in the two $\sigma_x$-sectors). $L_{\pm}$ denote two ground states with fractional fermions at the left edge. Schematically symmetric/antisymmetric superposition of these ground
states can be used to get ground states with different spin parities. The fractional charge remains
at the left edge, and the spin parity is characterized by occupation/absence of Majorana fermions. The scenario also applies for the right
edge. Notice however the MZMs are of many-body type in nature, and actually can not be constructed in terms of mean field states. }
\label{fig:Majorana zero modes}
\end{figure}

\textit{Edge zero modes}
Because of the oscillating sign before $M$ in Eq.(\ref{mft}), one of the sectors (labeled by $\sigma_x=\pm 1$) must be topological while the
other is trivial. In the topological sector, two-fold degenerate edge zero modes carrying fractional charges appear, similarly to the case in
the SSH model. (The entire system is a direct product of both sectors and is therefore always topologically nontrivial.) On the other hand,
the order parameter $M$ can take two opposite signs. Thus the ground states are 4-fold degenerate\cite{note1}. To gain insight into the
topological nature of the edge zero modes, it suffices to consider, without loss of generality, the special case $t=|M|$, in which the mean
field chain contains, for a given $\sigma_x$-sector, disconnected dimers in the bulk, while zero modes would be completely localized at the
dangling edges, as schematically shown in Fig.\ref{fig:Majorana zero modes}. We denote the ground states with edge modes (which will be
referred to simply as edge modes) as $\ket{E_{\pm}}$, where $\pm$ indicates the $\sigma_x$-sector, and $E=L/R$ the left/right edge. Such
states do not have definite spin parity, but can be recombined to do so, $\ket{E_{\ua/\da}}=(\ket{E_{+}}\pm \ket{E_{-}})/\sqrt{2}$. It is
clear that $E_{\ua/\da}$ carries definite $S_z=\pm 1/2$ and therefore they must differ in spin parity.

The charge density $\rho_E(j) =\bra{E_\sigma} c_j^\dag\sigma_0 c_j\ket{E_\sigma}$ in a state $\ket{E_\sigma}$ can be calculated
straightforwardly by inspection of Fig.\ref{fig:Majorana zero modes}. (The result is independent of $\sigma=\ua,\da$.) For example,
$\rho_L(1)=3/2$, $\rho_L(N) = 1/2$ and $\rho_L(j\neq 1,N) = 1$, while $\rho_R(1)=1/2$, $\rho_R(N) = 3/2$ and $\rho_R(j\neq 1,N) = 1$. Here
$N$ is the number of sites on the open chain.
Thus the edge modes can be probed by measuring the excess charge on the edges.

We now look into the spin property more closely. MZMs have to set in for the ground states with definite spin
parity $P_S=\pm 1$. They can not be constructed directly from the edge zero modes discussed so far, since they are essentially of
many-body type in nature. We may, however, construct many-body type MZMs at least formally~\cite{Ortiz-MF}: with
$d^\dagger_{E}=\ket{E_\da}\bra{E_\ua}$, Majorana operators can be defined as $\gamma_{E}=d^\dagger_{E}+d_{E}$ and
$\eta_{E}=-i(d_{E}-d^\dagger_{E})$\cite{note2}. When acting on a spin-parity definite many-body ground state, the Majorana operators switches
the spin polarity, and is exactly what can be used to classify the $\ZZ_2$ topology in the spin sector.

Even though an exact MZM wave function is unavailable at this stage, we may gain insights by inspecting the matrix element of spin-flipping
operator (which changes the spin parity) between the ground states, $s_E(j)=\bra{E_\ua} c_j^\dag\sigma_x c_j \ket{E_\da}$ for
$E=L/R$~\cite{manybody-MF}. By direct calculations, we find $s_L(1) =\frac{1}{2}$, $s_L(N) =-\frac{1}{2}$, $s_R(1) =-\frac{1}{2}$, $s_R(N)
=\frac{1}{2}$ and $s_{L/R}(j)=0$ elsewhere. The peaks in $s_E(j)$ imply the MZMs are also bound to the edges.

Given the existence of MZMs in the spin sector discussed above and the four-fold degeneracy in the ground state manifold, it appears plausible
to rearrange the four zero modes into a set of four MZMs spanned by, e.g., $\{\gamma,\eta\}\otimes\{\psi_L, \psi_R\}$, with $\{\gamma,\eta\}$
describing topological degeneracy due to the SPT fractionalization of $\ZZ_2$ spin parity, and $\{\psi_L, \psi_R\}$ describing topological
degeneracy due to the SPT fractionalization of inversion symmetry. In a cartoon picture, the edge zero modes can be understood as spin MZMs
(with amplitudes on both edges) decorated by a fractional charge (at one of the edges).

The product structure of edge zero modes enables unconventional braiding properties. The braiding may be applied either in the spin or charge
sector. For example, the unitary operator $T_1 =e^{\frac{\pi}{4}\gamma\eta}$ braids $\gamma$ and $\eta$~\cite{Ivanov,Alicea}, and
$T_2=e^{i\pi(F_R^\dagger F_R+F_L^\dagger F_L-F_L^\dagger F_R-F_R^\dagger F_L)}$ exchanges two complex fermions $F_L^\dag$ and $F_R^\dag$,
which can be constructed from the four MZMs~\cite{Loss-FF}. The braiding may be achieved by tuning $W_1$ and $W_2$ in a T-junction
set-up~\cite{Alicea}.  More interestingly, one may braid all of the four MZMs, $\gamma_L\rightarrow \eta_R, \ \eta_R\rightarrow \gamma_L,\
\eta_L\rightarrow -\gamma_R, \ \gamma_R\rightarrow-\eta_L$ with the unitary operator $U=T_1 T_2$.\\

\textit{Bosonization and field theoretical description}
We now go beyond mean field theory to gain further understanding of the edge zero modes. The low energy physics is most reliably captured by
the bosonized field theory. Following the standard procedure~\cite{Giamarchi}, we obtain an effective Hamiltonian $H=H_c+H_s + H_{m}$, with
\begin{eqnarray}
&&H_c = H_{0,c}+\frac{2g_c}{(2\pi a)^2}\int dx ~\cos(\sqrt{8}\phi_c),\nonumber \\
&&H_s = H_{0,s}+\frac{2g_s}{(2\pi a)^2}\int dx ~\cos(\sqrt{8}\phi_s) \nonumber \\
&&\ \ \ \ \ \ \ +\frac{2h_s}{(2\pi a)^2}\int dx ~\cos(\sqrt{8}\theta_s),\nonumber\\
&&H_m=\frac{2h_{m}}{(2\pi a)^2}\int dx~\cos(\sqrt{8}\phi_c) \cos(\sqrt{8}\theta_s).
\label{lattice-bosonized}
\end{eqnarray}
Here $H_{0,\nu}=\frac{1}{2}v_\nu\int dx[K_\nu(\partial_x\theta_\nu)^2+K_\nu^{-1}(\partial_x\phi_\nu)^2]$ for $\nu=c,s$, $\phi_{c/s}$ is the
bosonic field describing the charge/spin excitations with velocity $v_{c/s}$, $\frac{1}{\pi}\partial_x\theta_\nu$ is the conjugate momentum
of $\phi_\nu$, and $a$ is the lattice spacing. The Luttinger parameters are given by $K_c=(1+\frac{2W_1}{\pi v_F})^{-\frac{1}{2}}$ and
$K_s=(1-\frac{2W_1}{\pi v_F})^{-\frac{1}{2}}$. The mass parameters are $g_c=2W_1$, $g_s=2W_1$, $h_s=-2W_2$ and  $h_m=2W_2$. We notice that
under the inversion symmetry $P_I$ and spin parity $P_S$, the fields transform as $P_I^\dagger\phi_c(x)P_I=-\phi_c(-x)$ and
$P_S^\dagger\theta_s(x)P_S=\theta_s(x)+\frac{\pi}{\sqrt{2}}$ (see the Appendix~\ref{appendix-lattice-symmetries} for more details).

After bosonization, the system would be manifestly spin-charge separated if the mixing part $H_m$ were absent. In fact, the mass dimension of
$h_m$ is $2-2K_c-2K_s^{-1}$ (see the Appendix~\ref{appendix-lattice-renormalization}), being negative in the weak coupling limit where $K_{c,s}\sim 1$. Therefore we drop $H_m$ for
a moment, and will come back to its effect shortly. This enables us to address topological phases in the two sectors separately. Since
$K_c<1$ for $W_1>0$, $g_c$ is relevant and opens a charge gap. Meanwhile, $K_s<1$ so $g_s$ is irrelevant, but $h_s$ becomes relevant and also
opens a gap.  Thus $g_c$ and $h_s$ will flow to strong-coupling under RG. Since in our case $g_c>0$ and $h_s<0$, semiclassically the ground
state is characterized by $\phi_c= \frac{\pi}{\sqrt{8}} \ {\rm or}\  \frac{3\pi}{\sqrt{8}}$, and $\theta_s=0 \ {\rm or}\
\frac{\pi}{\sqrt{2}}$. The ground state is clearly 4-fold degenerate. Interestingly, $H_m$ is a win-win coupling that gains energy from both
fields in the above ground state configurations. It therefore enhances the stability of such ground states without spoiling the ground state
degeneracy. (A more detailed RG analysis of $H_m$ can be found in the Appendix~\ref{appendix-lattice-renormalization}.) Irrespective of $H_m$, the ground state develops
spin-charge separation, in the sense that the charge sector describes a bond insulator (or Pierls insulator), while the spin
sector describes the dual of spin-density-wave (a spin superfluid in a loose sense).

We notice that the mean field bSDW discussed above can be translated into $M\sim\avg{\sin(\sqrt{2}\phi_c)\cos(\sqrt{2}\theta_s)}$, which is
finite and may pick up two opposite signs in the above semi-classical ground states. This verifies the leading ordering tendency identified
by FRG. Moreover, as in the mean field theory case, the semiclassical state in the spin sector, say $\ket{\theta_s}$, mixes spin parities. We
can fix the parity by symmetric/antisymmetric recombination, $\ket{e/o}\equiv
\frac{1}{\sqrt{2}}(\ket{\theta_s=0}\pm\ket{\theta_s=\frac{\pi}{\sqrt{2}}})$. We find $\ket{e/o}$ is even/odd under $P_S$, with the
understanding that $\theta_s+\sqrt{2}\pi$ is equivalent to $\theta_s$. The degeneracy of ground states with respect to spin parities implies
that the spin sector may be mapped to a topological superfluid, as already allured to. To see further how this comes about,
we consider an enlightening case, the Luther-Emery point\cite{Luther-Emery} $(K_c,K_s)=(\frac{1}{2},2)$, at which the bosonic Hamiltonian
(dropping the $H_m$ part) can be exactly refermionized as
\begin{eqnarray}
 && H_s=\int dx \left\{ \chi_{s}^\dagger (-iv_s\partial_x\tau_3) \chi_{s}-
\frac{h_s}{2} [\chi_{s}^\dagger \tau_2(\chi_{s}^\dagger)^t+{\rm h.c.}]\right\}, \nonumber\\
&& H_c=\int dx (\chi_c^\dagger (-iv_c\partial_x \tau_3 - g_c\tau_2)\chi_{c}.
\label{eq:LutherEmery}
\end{eqnarray}
Here $\chi_{\nu}=(\chi_{\nu}^{R},\chi_{\nu}^{L})^t$ is a two-component spinor of chiral fermions in the $\nu=c/s$ channel, $\tau_{2,3}$ are
Pauli matrices in the chiral basis, and we dropped the irrelevant $g_s$-term for brevity. We observe that $H_{c}$ ($H_s$) is exactly
equivalent to the continuum limit of the SSH model (Kitaev model of 1D $p$-wave superconductor), with fractional fermions (Majorana zero
modes) at the edges. (In the Appendix~\ref{appendix-lattice-refermionization} we show that the effect of $H_m$ on top of the Luther-Emery Hamiltonian merely enhances the stability
of the edge modes.) The topological features, although obtained at a special point, are expected to hold as long as the gaps remains finite
in the bulk.

We remark that the spinor fields $\chi_\nu$'s are not simply related to the fundamental fields $c_j$, but should be viewed as
solitons in the $\phi_\nu$ fields\cite{Giamarchi}. Along this line, we find the fractional fermion modes can be viewed as kinks
in $\phi_c$, while the MZMs can be viewed as kinks in both $\phi_s$ and $\theta_s$ fields (see Appendix~\ref{appendix-lattice-kinks}).

The bosonic field theory corroborates our earlier analysis for the simultaneous presence of MZMs (in the spin sector) and fractional fermion
modes (in the charge sector). This is remarkable since they are two essentially different types of topological edge modes.
The spin-charge separation in the ground states makes it clear that the MZMs in our case are indeed of many-body type in nature.\\

\textit{Haldane phase in spin-1 chain}
We now illustrate that the well-known Haldane phase in spin-1 chain~\cite{Haldane-phase} may also be understood in terms of spin-charge
separated edge zero modes in an equivalent fermionic model. We consider the spin-1 XXZ chain described by the Hamiltonian
\begin{equation}
H_{XXZ} = \sum_{j}^{}J(S_j^x S_{j+1}^x+S_j^y S_{j+1}^y)+ J_z\sum_j S_j^z S_{j+1}^z ,
\label{model-xxz}
\end{equation}
where $J>0$ and $J_z>0$.  In a regime of parameters, including the isotropic point $J=J_z$, the ground state of the spin-1 system is 4-fold
degenerate, characterized by deconfined spinons $S_z=\pm 1/2$ at the edges~\cite{Berg-rise}.
Under a generalized Jordan-Wigner transformation~\cite{GJW}, the spin-1 XXZ chain is mapped to a spin-1/2 fermion model~\cite{t-jz,GJW}
\begin{eqnarray}
{\cal H}=&& J \sum_{j\sigma} ( \bar{c}^\dagger_{j \sigma} \bar{c}_{j+1\sigma} +\bar{c}^\dagger_{j\sigma} \bar{c}^\dagger_{j+1 \bar{\sigma}} +
{\rm
h.c.})\nonumber\\ &&+ 4J_z\sum_j S_j^z S_{j+1}^z,
\label{xxz-fermionized}
\end{eqnarray}
where $\bar{c}_{j \sigma}=c_{j\sigma}(1-n_{j\bar{\sigma}})$ is the fermion operator subject to no double occupancy,
$n_{j\sigma}=c_{j\sigma}^\dag c_{j\sigma}$ and $S_j^z=(n_{j\ua}-n_{j\da})/2$. The bosonization can be performed by softening the hard
constraint, $\bar{c}_{j\sigma}\ra c_{j\sigma}(1-\epsilon n_{j\bar{\sigma}})$, in the spirit of adiabatic continuality from $0<\epsilon<1$ to
$\epsilon=1$\cite{wu}. Without the p-wave triplet pairing terms in ${\cal H}$, the model becomes the so-called $t-J_z$ model, which is known to be gapless in
the charge sector. In the presence of the pairing terms, however, the charge and spin sectors are mixed so that all excitations are gapped in
the bulk, as in the spin-1 model. Interestingly, the ground states of ${\cal H}$ are also four-fold degenerate, and the roles of spin and
charge with regard to the SPT fractionalization are exchanged (see Appendix~\ref{appendix-Haldane}). This is not surprising because ${\cal
H}$ has a $\ZZ_2$ fermion parity. At the Luther-Emery fixed point, the refermionized Hamiltonian would be essentially equivalent to
Eq.(\ref{eq:LutherEmery}) upon the exchange $s \leftrightarrow c$. Now the charge sector describes topological
`superconductivity', while the spin sector depicts a spin-gapped insulator. The four-fold ground state degeneracy can be characterized by
Majorana zero modes in the charge sector decorated by fractional fermions in the spin sector. For comparison, it is the spin that is bound to
MZMs in the DIII-class topological superconductor. Thus the ${\cal H}$ model describes a new type of 1D topological superconductor.\\

\textit{Summary}
We have demonstrated that spin-charge separated Majorana modes (in the spin sector) and fractional fermions (in the charge sector) can present
simultaneously in a 1D chain following from SPT fractionalizations of inversion symmetry and spin parity. We have also
offered an alternative understanding of the Haldane phase in terms of  spin-charge separated edge zero modes. The lattice model we proposed
and the novel properties may be simulated and probed by cold atoms in optical lattices.\\

\begin{acknowledgments}
We thank  Y. X. Zhao and Y. Chen for helpful discussions. This work was supported by the GRF of Hong Kong ( HKU173051/14P $\&$ HKU 173055/15P), the URC fund of HKU, and NSFC (under grant No.11574134).
\end{acknowledgments}

\appendix

\section{Singular-mode functional renormalization group}
\label{sm-frg}
In the presence of competing orders, FRG is advantageous to judge the leading ordering tendency at low energy scales. Consider the
interaction hamiltonian \begin{equation}
H_I=\frac{1}{(2!)^2} c_1^\dagger c_2^\dagger V_{1234} c_3 c_4.\end{equation}
Henceforth the numerical index labels momentum $k$ and spin $\sigma$, and we leave implicit the overall momentum conservation
$k_1+k_2=k_3+k_4$. The interaction vertex $V_{1234}$ is fully anti-symmetrized with respect to $(1,2)$, and to $(3,4)$. For brevity,
summation over repeated indices is implied unless declared otherwise. (The normalization constant in the summation over momentum is absorbed
by assuming unit length of the chain.) The idea of FRG is to get the effective one-particle-irreducible interaction vertex function
$\Gamma_{1234}$ for fermions whose energy/frequency is above a scale $\Lambda$. (Thus $\Gamma$ is $\Lambda$-dependent.) Equivalently, such a
vertex function may be understood as a generalized pseudo-potential for fermions whose energy/frequency is below $\Lambda$. Starting from the
bare vertex $V_{1234}$ at $\Lambda\gg 1$, the contributions to $\partial \Gamma/\partial\Lambda$ with decreasing $\Lambda$ is given by, with
full fermion antisymmetry,
\begin{eqnarray}
\frac{\partial}{\partial\Lambda}\Gamma_{1234} =&& -\frac{1}{2}\Gamma_{1265}\chi^{pp}_{56}\Gamma_{5634} +
\Gamma_{1635}\chi^{ph}_{56}\Gamma_{5264} \nonumber \\ &&- \Gamma_{1564}\chi^{ph}_{56}\Gamma_{6235},
\end{eqnarray}
where
\begin{eqnarray}
&&\chi^{pp}_{12} = \frac{\partial}{\partial\Lambda}\int\frac{d\omega}{2\pi}G_1(\omega)G_2(-\omega)\theta(|\omega|-\Lambda),\\
&&\chi^{ph}_{12} = -\frac{\partial}{\partial\Lambda}\int\frac{d\omega}{2\pi}G_1(\omega)G_2(\omega)\theta(|\omega|-\Lambda),
\end{eqnarray}
are differential susceptibilities in the particle-particle (pp) and particle-hole (ph) channels, and $G_1(\omega)$ is the
normal state Green's function at Matsubara frequency $\omega$ for the single-particle Bloch state labeled by $1$. If the momentum dependence
in $\Gamma_{1234}$ is projected to the Fermi points, FRG becomes equivalent to the $g$-ology RG and the so-called patch-FRG \cite{patchfrg}
(if applied in the 1D case). Because of the limitation in momentum resolution, such RG schemes are known to be insufficient to describe
non-local order parameters in 1D~\cite{Honerkamp-so(5)}. Since retaining the full momentum dependence otherwise in $\Gamma_{1234}$ is an
insurmountable task, we need a suitable truncation scheme to keep the most important (or potentially singular) part of $\Gamma_{1234}$. This
is achieved in the singular-mode FRG (SM-FRG)\cite{smfrg}. In short, the potentially singular part of $\Gamma$ can be expanded in terms of
scattering matrices between finite-ranged (up to a truncation length $r_c$) fermion bilinears in the pp and ph channels. (Notice that the
scattering distance between fermion bilinears is free from truncation.) This truncation scheme is asymptotically exact in the limit of
$r_c\ra \infty$, and in practice a finite $r_c$ is sufficient to capture any order parameters that can be defined on site up to on bonds of
length $r_c$. In our case we find the result converges already at $r_c=2$. The rational behind the success of this truncation scheme is the
fact that order parameters following from collective modes are in general short-ranged in internal structure. The technical details can be
found elsewhere\cite{smfrg}. Here we merely quote that from $\Gamma_{1234}$ we can extract the effective interaction $V^{pp}$ and $V^{ph}$ in
the pp and ph channels, respectively,
\begin{equation}
V^{pp}_{(1|2)(4|3)}=\Gamma_{1234}, \ V^{ph}_{(1|3)(4|2)}=-\Gamma_{1234},
\end{equation}
where $(1|2)$ is a combined index for a fermion bilinear. The fact that both interactions are extracted from the same vertex function means
that all channels are treated on equal footing. The interaction matrices can be decomposed into eigen modes as, with explicit momentum-spin
indices, and for a given collective momentum $q$,
\begin{eqnarray}
V^{pp}_{(k+q,\alpha|\bar{k},\beta)(k'+q,\gamma|\bar{k}',\delta)} = \sum_m S_m^{pp} \phi_m^{\alpha\beta}(k)
(\phi_m^\dag)^{\gamma\delta}(k'),\nonumber\\
V^{ph}_{(k+q,\alpha|k,\beta)(k'+q,\gamma|k',\delta)} = \sum_m S_m^{ph} \psi_m^{\alpha\beta}(k) (\psi_m^\dag)^{\gamma\delta}(k'),\nonumber
\end{eqnarray}
where $\bar{k}=-k$, $m$ labels the eigen mode, $S_m$ is the eigenvalue, and $\phi_m$ or $\psi_m$ is the (matrix) eigen function. Up to
symmetry-dictated degeneracy, the most diverging (versus decreasing running scale $\Lambda$) and attractive eigen mode indicates the
instability channel and the associated eigen function describes the emerging order parameter.

\begin{figure}
\includegraphics[width=10cm]{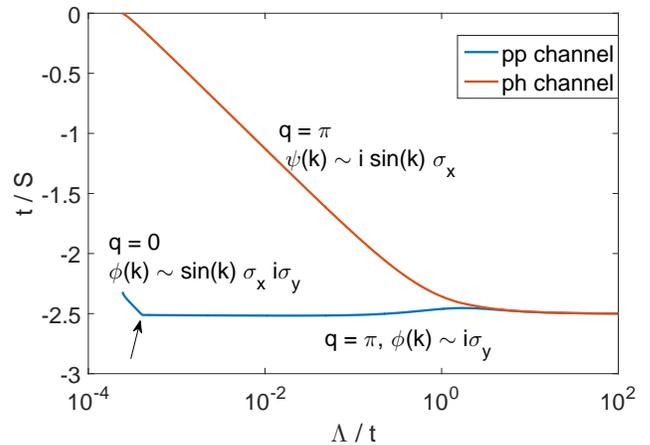}
\caption{FRG flow of the leading attractive eigenvalues $S$ (plotted as $1/S$ for a better view) in the pp (blue) and ph (read) channels for
$W_1=W_2=0.2t$. The associated collective momentum $q$ and (matrix) form factor are also given. Notice that they may evolve during the flow
to low energy scales. For example, in the pp channel there is a level crossing (arrow) from $q=\pi$ at higher energy scale to $q=0$ at lower
energy scale.} \label{frg}
\end{figure}

Fig.\ref{frg} shows the flow of the leading attractive eigenvalues of the effective interactions in the pp and ph channels versus the
decreasing energy scale $\Lambda$ for the bare parameters $W_1=W_2=0.2t$. The ph channel (red line) is clearly dominant. The collective
momentum $q=\pi$ and the form factor $\psi(k)\sim i\sin k \sigma_x$ describe exactly the structure of bSDW, $M\sim \sum_k \avg{
c^\dag_{k+\pi}i\sin k \sigma_x c_k\>}$. The pp channel (blue line) is weak. At higher energy scales, the leading mode in this channel would
describe an sPDW with collective momentum $q=\pi$ and form factor $\phi(k)\sim i\sigma_y$. At lower scales, it switches to a uniform triplet
pairing with collective momentum $q=0$ and form factor $\phi(k) = \sin k \sigma_x i\sigma_y$. These modes are mentioned in the main text. The
weakness of the pairing channel results from interference between umklapp scattering and Cooper scattering.

We have performed systematic calculations in the regime $0<W_1\leq0.3t$ and $0<W_2\leq 0.3t$. We find that bSDW is the leading instability
for $W_1/W_2 > 1/8$, while the usual site-local SDW (with Neel moment along $\hat{x}$) becomes the leading instability for $W_1/W_2<1/8$. In
the secondary pp channel, the leading mode (at the divergence scale of the ph channel) corresponds to sPDW for $W_1/W_2>1$, and to triplet
pairing for $W_1/W_2<1$. We conclude that the mean field theory in the main text is valid as long as $W_1/W_2>1/8$.

\section{Bosonization and field theory of the lattice model}
\label{appendix-lattice-bosonization}
\subsection{Bosonization}
In the following we describe the technical details in the bosonization of fermion model (\ref{model-lattice}). In the low energy and long
wavelength limit, we have
\begin{equation}
\frac{c_j}{\sqrt{a}}\rightarrow\psi_{\sigma}(x)=\psi_{R\sigma}(x)e^{ik_Fx}+\psi_{L\sigma}(x)e^{-ik_Fx}
\end{equation}
where $\psi_{\epsilon\sigma}(x)$ (with $\epsilon=\pm$) describes right/left moving chiral fermions, $a$ is the lattice spacing  and
$k_F=\frac{\pi}{2}$ is the Fermi momentum. Using standard bosonization techniques\cite{Giamarchi}, the chiral fermions can be expressed
through boson fields as,
\begin{equation}
    \psi_{\epsilon\sigma}(x)=\frac{U_{\epsilon\sigma}}{\sqrt{2\pi a}}e^{i[\theta_\sigma(x)-\epsilon\phi_\sigma(x)]}
\end{equation}
where $\phi_\sigma(x)$ and $\theta_\sigma(x)$ are boson fields subject to $[\phi_\sigma(x),
\frac{1}{\pi}\partial_x\theta_{\sigma'}(x')]=-i\delta(x-x')\delta_{\sigma\sigma'}$, and $U_{\epsilon\sigma}$ is the Klein factor insuring
anticommuting relation between fermions of different species. To reveal the charge and spin degree of freedom in this system, one turns to
the new basis
\begin{eqnarray}
 \phi_{c/s}=\frac{1}{\sqrt{2}}(\phi_\ua\pm\phi_\da), \nonumber \\
 \theta_{c/s}=\frac{1}{\sqrt{2}}(\theta_\ua\pm\theta_\da).
\end{eqnarray}
with $c/s$ denoting charge/spin.

We first bosonize the noninteracting part of $H$ to get $H_0^\nu=\frac{v_F}{2}\int
dx[(\partial_x\theta_\nu(x))^2+(\partial_x\phi_\nu(x))^2]$, where $\nu=c/s$.  We proceed to bosonize the interactions. We observe that
\begin{eqnarray}
&& \psi_\ua^\dagger(x)\psi_\da^\dagger(x)=
\psi_{R\ua}^\dagger\psi_{L\da}^\dagger+\psi_{L\ua}^\dagger\psi_{R\da}^\dagger  \nonumber \\
&&\ \ \ \ \ +\psi_{R\ua}^\dagger\psi_{R\da}^\dagger e^{-2i k_Fx}+\psi_{L\ua}^\dagger\psi_{L\da}^\dagger e^{+2i k_Fx}\nonumber \\
&&\psi_\da(x+a)\psi_\ua(x+a)=\psi_{R\da}\psi_{L\ua}+\psi_{L\da}\psi_{R\ua}\nonumber \\
&&\ \ +\psi_{R\da}\psi_{R\ua} e^{-2i k_F(x+a)}+\psi_{L\da}\psi_{L\ua} e^{+2i k_F(x+a)}.
\end{eqnarray}
Plugging into the $W_1$-term we get
\begin{eqnarray}
\frac{2W_1}{(2\pi a)^2}\{[2a^2(\partial_x\phi_c)^2-2a^2(\partial_x\phi_s)^2]\nonumber\\ +2\cos\sqrt{8}\phi_c+2\cos\sqrt{8}\phi_s)\}.
\end{eqnarray}
To determine the sign of mass terms $\cos\sqrt{8}\phi_c$ and $\cos\sqrt{8}\phi_s$,  the ordering of the Klein factors matters. Henceforth we
use the convention $U_{R\ua}U_{L\ua} U_{L\da}U_{R\da}  = 1$. On the other hand, we observe that
\begin{eqnarray}
&\psi_\ua^\dagger(x)\psi_\da(x)=\psi_{R\ua}^\dagger\psi_{R\da}+\psi_{L\ua}^\dagger\psi_{L\da} \nonumber \\
&+\psi_{R\ua}^\dagger\psi_{L\da} e^{-2i k_Fx}+\psi_{L\ua}^\dagger\psi_{R\da} e^{+2i k_Fx} \nonumber \\
&\psi_\ua^\dagger(x+a)\psi_\da(x+a)= \psi_{R\ua}^\dagger\psi_{R\da}+\psi_{L\ua}^\dagger\psi_{L\da}\nonumber \\
&+\psi_{R\ua}^\dagger\psi_{L\da} e^{-2i k_F(x+a)}+\psi_{L\ua}^\dagger\psi_{R\da} e^{+2i k_F(x+a)}.
\end{eqnarray}
Substitution into the $W_2$-term yields
\begin{eqnarray}
-\frac{2W_2}{(2\pi a)^2}\cos\sqrt{8}\theta_s+\frac{4W_2}{(2\pi a)^2}\cos\sqrt{8}\phi_c \cos\sqrt{8}\theta_s.
\end{eqnarray}
Here we omitted the term $\cos\sqrt{8}\phi_s \cos\sqrt{8}\theta_s$ which is always irrelevant
since $\phi_s$ and $\theta_s$ are dual variables that can not be pinned simultaneously. Collecting everything together
we end up with Eq.(\ref{lattice-bosonized}).

\subsection{Bosonic renormalization group}
\label{appendix-lattice-renormalization}

\begin{figure}
\centering
\includegraphics[width=9cm]{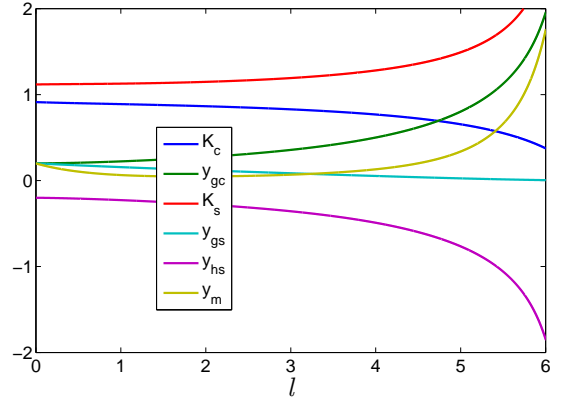}
\caption{RG flow of the coupling and Luttinger parameters. The initial condition is set as $y_{gc}=y_{hc}=-y_{hs}=y_m=0.2$, $
K_c=(1+0.2)^{-0.5}$ and $K_s=(1-0.2)^{-0.5}$. Notice that $y_m$ initially decreases but is eventually driven to strong coupling. See the text for details.}\label{fig:RG flow}
\end{figure}

Now we apply RG to analyze the bosonized $H$.
We define the dimensionless coupling parameters as $y_{gc}=\frac{g_c}{\pi v_c}$, $y_{gs}=\frac{g_s}{\pi v_s}$,  $y_{hs}=\frac{h_s}{\pi v_s}$,
and $y_{m}=\frac{h_m}{\pi v_s}$. Under RG, the coupling parameters and the Luttinger parameters flow as follows, up to the second order in
$y$'s,
\begin{eqnarray}
&&\frac{dy_{gc}}{dl}=(2-2K_c)y_{gc}-(1-2K_s^{-1})y_{m}y_{hs},  \nonumber \\
&&\frac{dK_c}{dl}=-\frac{1}{2}K_c^2(y_{gc}^2+y_{m}^2),
\end{eqnarray}
for the charge sector, and
\begin{eqnarray}
&&\frac{dy_{gs}}{dl}=(2-2K_s)y_{gs}, \nonumber \\
&&\frac{dy_{hs}}{dl}=(2-2K_s^{-1})y_{hs} -(1-2K_c)y_{gc}y_m, \nonumber \\
&&\frac{dK_s}{dl} =\frac{1}{2}(y_{hs}^2+y_m^2-K_s^2y_{gs}^2),
\end{eqnarray}
for the spin sector, and
\begin{eqnarray}
\frac{dy_{m}}{dl}=(2-2K_c-2K_s^{-1})y_{m} -y_{gc}y_{hs}
\end{eqnarray}
for the mixing term in $H_m$. Here $l=\ln \frac{a'}{a}$ is the RG parameter. As stated in the main text, we choose $W_1>0$ and $W_2>0$ so
that initially $y_{gc}>0$, $y_{hc}>0$, $K_c<1$, $K_s>1$ and $y_{hs}=-y_m<0$. Numeral solution of the RG equations presented in
Fig.\ref{fig:RG flow} reveals that eventually $g_c$, $h_s$ and $h_m$ are relevant, while $g_s$ is irrelevant.
Even though the mass dimension of $h_m$ is $2-2K_c-2K_s^{-1}<0$ initially, it becomes relevant at large $l$ (or small energy scale) because
of a source term in the flow equation, $-y_{gc}y_{hs}$, which is positive in our case and drives $y_m$ to strong coupling.

\subsection{Refermionization}
\label{appendix-lattice-refermionization}

To gain further insights into the gapped phase, we study the model at the Luther-Emery point. We first ignore the mixing term
$H_m$. This enables us to address edge zero modes from spin and charge channels separately. In the charge sector, the Luther-Emery point is
$K_c=\frac{1}{2}$. To refermionize $H_c$, we first rescale the bosonic fields as $\frac{\phi_c}{\sqrt{K_c}}\rightarrow \phi_c$, $\sqrt{K_c}
\theta_c \rightarrow \theta_c$.  Introducing new chiral fermionic operators,
\begin{eqnarray}
\chi^R_{c}=\frac{U^R_{c}}{\sqrt{2\pi a}}e^{-i(\phi_c-\theta_c)}, ~~\chi^L_{c}=\frac{U^L_{c}}{\sqrt{2\pi a}}e^{-i(\phi_c-\theta_c)}
\end{eqnarray}
we get
\begin{eqnarray}
H_c=\int dx \chi_c^\dag( -i v_c\tau_3\partial_x - g_c\tau_2)\chi_c,
\label{refermion_Hc}
\end{eqnarray}
where $\chi_c=(\chi_c^R,\chi_c^L)$ is a spinor and $\tau_{2,3}$ are Pauli matrices in the chiral basis.
In this form $H_c$ corresponds to the continuum limit of SSH model. The system supports zero modes of fractional fermions at two ends,
created by the following field operators
\begin{eqnarray}
&& F_L^\dagger=\sqrt{\frac{g_c}{v_c}}\int_0^N dx(\chi_{c}^R+\chi_{c}^L)^\dagger e^{-\frac{g_c}{v_c}x}, \nonumber \\
&& F_R^\dagger=\sqrt{\frac{g_c}{v_c}}\int_0^N dx(\chi_{c}^R-\chi_{c}^L)^\dagger e^{-\frac{g_c}{v_c}(N-x)},
\label{edge_FF}
\end{eqnarray}
where $N$ is the length of the chain.

In the spin sector, the Luther-Emery point is at $K_s=2$, and we want to refermionize $H_s$, dropping the irrelevant term $g_s$ for brevity.
Similarly to the case of charge sector, we rescale the bosonic fields as $\frac{\phi_s}{\sqrt{K_s}}\rightarrow \phi_s$, $\sqrt{K_s} \theta_s
\rightarrow \theta_s $ and introduce new chiral fermionic operators,
\begin{eqnarray}
\chi_{s}^R=\frac{U_{s}^R}{\sqrt{2\pi a}}e^{-i(\phi_s-\theta_s)}, ~~\chi_{s}^L=\frac{U_{s}^L}{\sqrt{2\pi a}}e^{-i(\phi_s+\theta_s)},
\end{eqnarray}
and we end up with
\begin{eqnarray}
H_s=\int dx \left[ \chi_s^\dag( -i v_s\tau_3\partial_x)\chi_s - \frac{h_s}{2}( \Delta_s^\dag +\Delta_s)\right],
\end{eqnarray}
where $\chi_s=(\chi_s^R,\chi_s^L)$ is the spinor in the chiral basis, and $\Delta_s^\dag =\chi_{s}^\dagger \tau_2(\chi_s^\dag)^t$ is a
triplet pairing operator in the spin sector. Clearly $H_s$ in the above form corresponds to the continuum limit of Kitaev model of one
dimensional $p$-wave superconductor. The two Majorana zero modes are obtained as
\begin{eqnarray}
&&\gamma_L=\sqrt{\frac{|h_s|}{2v_s}}\int_0^N dx ~i(\chi_{s}^R +\chi_{s}^L-{\rm h.c.} )e^{-\frac{|h_s|}{v_s}x}, \nonumber \\
&&\gamma_R=\sqrt{\frac{|h_s|}{2v_s}}\int_0^N dx(\chi_{s}^R+\chi_{s}^L + {\rm h.c.}) e^{-\frac{|h_s|}{v_s}(N-x)}.
\label{edge_MF} \nonumber
\end{eqnarray}

Now we consider the effect of the mixing term $H_m$. We observe that the pinning of
$\phi_c$ and $\theta_s$ also minimizes $H_m$. Therefore we expect the topological properties of this system is not changed by $H_m$, which
can be written as, at the Luther-Emery point,
\begin{eqnarray}
H_m=\frac{h_m}{2}\int dx ~\chi_{c}^\dag \tau_2\chi_c (\Delta_s^\dag +\Delta_s). \nonumber
\end{eqnarray}
In the presence of $H_m$, $[H, F_{L/R}^\dagger]\neq 0$ and $[H,\gamma_{L/R}]\neq 0$, meaning that the edge modes obtained earlier are no
longer zero-energy eigen modes. However, we may find new edge zero modes in a mean-field approximation,
\begin{equation}
H_m \sim  \int dx \left[h_m \delta_s \chi_c^\dag\tau_2\chi_c + h_m \delta_c (\Delta_s^\dag +\Delta_s)/2\right],
\end{equation}
with $\delta_s=\avg{\Delta_s^\dag+\Delta_s}/2$ and $\delta_c=\avg{\chi_c^\dag\tau_2\chi_c}$. This merely modifies the mass terms as $
\bar{g}_c=g_c-h_m \delta_s$ and $\bar{h}_s=h_s-h_m \delta_c$.
Given $g_c>0$ ($h_s<0$) in our case, $H_c$ ($H_s$) fixes $\delta_c>0$ ($\delta_s <0$), so that $\bar{g}_c>g_c>0$ and $\bar{h}_s<h_s<0$. Thus
edge zero modes can be reconstructed, and the only change is the enhancement of energy gaps in the bulk and reduction in the penetration
depth of the edge modes. This is a restatement that $H_m$ further stabilizes the edge zero modes.

\subsection{Edge zero modes as kinks }
\label{appendix-lattice-kinks}
The Edge zero modes can also be regarded as kinks in bosonic fields.  Kinks corresponding to edge zero modes of fractional fermions and
Majorana fermions are fundamentally different, and thus we would like to discuss them separately.  Before doing so, we should first
understand what is the proper field theory for the vacuum.  We may consider the vacuum as a trivial insulator in both spin and charge
sectors, with the mass term $-V(\cos2\phi_\ua+\cos2\phi_\da)=-2V\cos\sqrt{2}\phi_c\cos\sqrt{2}\phi_s$. We assume $V>0$ so that $\phi_c$ and
$\phi_s$ are pinned at zero. In this scenario there is no spin-charge separation in vacuum. This will be very important in the discussion of
Majorana zero modes in the spin sector below.

Assume the quantum wire is bounded by  $x\in [0,N]$. Let us first consider edge zero modes of fractional fermions. Since $\phi_c$ is pinned
to $0$ in the vacuum and to $\frac{\pi}{\sqrt{2}}(n_{\phi_c}+\frac{1}{2})$ in the quantum
wire, where $n_{\phi_c}$ is integer operator, there must be a kink connecting the two phases across the boundary.
This kink is of minimal magnitude $\pm\frac{\pi}{\sqrt{8}}$, resulting a fractional fermion located at the boundary with  charge
\begin{equation}
\rho_{F}=\int dx \rho_c(x)=-\int dx\frac{\sqrt{2}}{\pi}\partial_x \phi_c=\mp \frac{1}{2}.
\end{equation}

Now we address the Majorana zero modes. In the quantum wire, we have $\theta_s=\frac{\pi}{\sqrt{2}}n_{\theta_s}$. In the vacuum,  bosonic
fields are pinned as $\phi_s^{(j)}=\sqrt{2} \pi n_{\phi_s}^{(j)}$, where $j=1$ ($j=2$) refers to the vacuum at $x<0$ ($x>N$). The vacuum and
the quantum wire can be connected by the kink operators\cite{Clarke-pf}, $\gamma_{j} \sim e^{i\pi(n_{\phi_s}^{(j)}+n_{\theta_s})}$. Note that
$n_{\phi_s}^{(j)}$ and $n_{\theta_s}$ are integer operators, and $[\phi_s(x),\theta_s(x')]=i\pi\varTheta(x-x')$, where $\varTheta$ is the Heaviside step function. This leads to
$[n_{{\phi}_s}^{(2)}, n_{\theta_s}]=\frac{i}{\pi}$ and $[n_{{\phi}_s}^{(1)}, n_{\theta_s}]=0$ (due to the step function in the above
commutator), and consequently $\gamma_{j}^2=1$ and $\{\gamma_{1},\gamma_{2}\}=0$, exactly the required algebra for Majorana operators. This
defines the MZMs on the two edges. Note the fundamental difference to the kink operators for the fractional fermions.

The above discussion clarifies how edge zero modes can be realized in spin and charge channels separately. The fact that the MZMs are kinks
in both $\phi_s$ and $\theta_s$ signifies the many-body nature in such modes.

\subsection{Symmetries}
\label{appendix-lattice-symmetries}
Here we consider how inversion symmetry $P_I$ and spin parity symmetry $P_S$ operator work on the bosonic fields.

The inversion operator $P_I$ acts on charge/spin density as $P_I^+\rho_{c/s}(x)P_I=\rho_{c/s}(-x)$, and on charge/spin current as
$P_I^+j_{c/s}(x)P_I=-j_{c/s}(-x)$. We recall that
\begin{eqnarray*}
&& \rho_{c}=-\frac{\sqrt{2}}{\pi}\partial_x\phi_{c},\ \ \ \ \ \ \  \rho_{s}=-\frac{1}{\sqrt{2}\pi}\partial_x\phi_{s}, \\
&& j_{c}=\frac{\sqrt{2}}{\pi}\partial_x\theta_{c},\ \ \ \ \ \ \ \ \ j_{s}=\frac{1}{\sqrt{2}\pi}\partial_x\theta_{s}.
\end{eqnarray*}
Thus we have $P_I^+\phi_{c/s}(x)P_I=-\phi_{c/s}(-x)$ and $P_I^+\theta_{c/s}(x)P_I=\theta_{c/s}(-x)$.
As a side remark, the inversion operator for the refermionized $H_c$ in the main text can be determined by requiring $P_I^\dag h_c(k)
P_I=h_c(-k)$ where $h_c(k)$ is the single-particle part of $H_c$ in the momentum space. A simple inspection reveals that $P_I=\tau_2$.

Now we turn to the spin parity $P_S\equiv(-1)^{\rho_s}=e^{i\pi\rho_s}$.  Using the relation $[\rho_s(x),\theta_s(x')]=i\delta(x-x')$, we find
$e^{i\pi\rho_s} \theta_s e^{-i\pi\rho_s}=\theta_s+\frac{\pi}{\sqrt{2}}$. Thus, the spin parity operator shifts the bosonic field $\theta_s$
by $\frac{\pi}{\sqrt{2}}$.  The spin parity provides an intrinsic particle-hole symmetry for the refermionized $H_s$.

In the presence of fermion parity symmetry, we may also define a fermion parity operator $P_c\equiv(-1)^{\rho_c}=e^{i\pi\rho_c}$. We observe
that $[\rho_c(x),\theta_c(x')]=2i\delta(x-x')$, thus $e^{i\pi\rho_c} \theta_c e^{-i\pi\rho_c}=\theta_c+\sqrt{2}\pi$. This is very different
to the case of $P_S$ and will be useful when we discuss topological superconductivity associated with fermion parity.

\section{Bosonization of $t-Jz$ model with $p$-wave like pairing}
\label{appendix-Haldane}
The spin-1 Haldane chain can be mapped to a spin-1/2 fermion system described by the Hamiltonian ${\cal H}={\cal H}_{t-J_z}+{\cal
H}_p$,\cite{t-jz,GJW} with
\begin{eqnarray}
&& {\cal H}_{t-J_z} = J \sum_{j\sigma} ( \bar{c}^\dagger_{j \sigma} \bar{c}_{j+1 \sigma} +{\rm h.c.}) + 4J_z\sum_{j}S_j^z
S_{j+1}^z,\nonumber\\
&& {\cal H}_p = J\sum_{j\sigma}(\bar{c}^\dagger_{j \sigma} \bar{c}^\dagger_{j+1 \bar{\sigma}} + {\rm
h.c.} ).
\label{xxz-fermionized}
\end{eqnarray}
Here $\bar{c}_{j \sigma}=c^\dagger_{j\sigma}(1-n_{j\bar{\sigma}})$ is the fermion operator subject to no double occupancy constraint,
$n_{j\sigma}=c_{j\sigma}^\dag c_{j\sigma}$ and $S_j^z=(n_{j\ua}-n_{j\da})/2$. (Notice that no constraint is needed in $S^z_j$). In the limit
of $J_z=J$, ${\cal H}$ is equivalent to the isotropic spin-1 Heissenberg model.   Without the triplet-pairing term in ${\cal H}_p$, ${\cal H}_{t-J_z}$  can be bosonized by
softening the hard constraint on the fermion operators, $\bar{c}_\sigma \ra c_\sigma (1-\epsilon n_{\bar{\sigma}})$, in the spirit of adiabatic
continuity from $0<\epsilon<1$ to $\epsilon=1$\cite{wu},
\begin{eqnarray}
{\cal H}_{t-J_z} \sim && \sum_{\nu=c,s}\frac{v_\nu}{2}\int \left[ K_\nu(\partial_x\theta_\nu)^2 +
K_\nu^{-1}(\partial_x\phi_\nu)^2\right]\nonumber\\
&& +\frac{2g_s}{(2\pi\alpha)^2}\int dx \cos(\sqrt{8}\phi_s).
\label{t-Jz-bosonized}
\end{eqnarray}
The parameters are given by
\begin{eqnarray*}
&&K_c=\frac{1}{\sqrt{1 + \frac{4\epsilon}{\pi}\cot \frac{\pi f}{2}}},\\
&&K_s=\frac{1}{\sqrt{1+\frac{4J_z}{\pi v_F} -\frac{4\epsilon}{\pi}\cot\frac{\pi f}{2}} },\\
&&g_s=v_F\left[8\epsilon\cot\frac{\pi f}{2}+\frac{4J_z}{v_F}\cos(\pi f) -\frac{2\epsilon^2}{\pi}\right],
\end{eqnarray*}
where $f=2/3$ is the average filling, $v_F=2J\sin(\pi f/2)$ is the fermi velocity in the otherwise free model. Notice that the fermion system
is hole doped, with equal probability of spin-up fermion, spin-down fermion, and hole occupancies (so that $f=2/3$ on average), in order for 
${\cal H}$ to be equivalent to the spin-1 model in its spin-disordered phase. On one hand, this weakens the Mottness as in the
doped $t-J$ model, making the constraint-softening better defined, on the other hand, the umklapp term drops out of the charge sector.
Therefore the charge sector of ${\cal H}_{t-J_z}$ is gapless, while the spin sector is gapped if $K_s<1$. This behavior is significantly
modified by ${\cal H}_p$, which can be bosonized as
\begin{eqnarray}
{\cal H}_p\sim \frac{2h_m}{2\pi a}\int dx \sin(\sqrt{2}\theta_c)\sin(\sqrt{2}\phi_s).
\end{eqnarray}
Consistent with previous convention for the Klein factors, we used $U_{R\ua}U_{L\da}=U_{L\ua}U_{R\da}=i$ (the sign before the imaginary
number is understood as a gauge choice),  and we found $h_m\sim -2J(1-\epsilon f)$, apart from other contributions to ${\cal H}_p$ that are immaterial 
for the following discussion. ${\cal H}_p$ in the above form mixes the spin and charge sectors, and generates a gap in
the charge sector if $K_c>1/4$. In this way all excitations are gapped in the bulk, as anticipated for the original spin-1 chain.

More interestingly, the ground states of ${\cal H}={\cal H}_{t-J_z}+{\cal H}_p$ are also four-fold degenerate, provided that $g_s>0$. Indeed,
$\phi_s$ is pinned at $\lambda \pi/\sqrt{8}$, with $\lambda=\pm 1$. These spin states are bond insulators, and carry the amzaing Haldane
string order\cite{nonlocal},
\begin{equation*}
\avg{O_s^z(i,j)} \sim \avg{\sin\sqrt{2}\phi_s(i)\sin\sqrt{2}\phi_s(j)} \neq 0,\end{equation*}
where $O_s^z(i,j)=S_{i}^z e^{i2\pi\sum_{l=i+1}^{j-1} S_{j}^z}S_{j}^z$. Corresponding to $\phi_s$,
in the charge sector $\theta_c=\lambda (\pi/\sqrt{8}+n\sqrt{2}\pi)$ with $n=0,1$. These charge states are exactly related by the charge
parity operator $P_c$ we highlighted above. In total, we get four-fold degeneracy in the ground state manifold.

The properties of the spin and charge sectors provide the ground for edge zero modes. There are fractional fermions in the spin sector by
fractionalization of the inversion symmetry, and there are Majorana modes in the charge sector by fractionalization of the fermion parity. To
have a better idea, we first approximate $\sin(\sqrt{2}\phi_s)$ by its value in one of the semi-classical ground states. This is doable since
at the Gaussian level, the fluctuations of $\sin(\sqrt{2}\phi_s)$ is four times smaller than that of $\cos(\sqrt{8}\phi_s)$. Under this
approximation the resulting model can be refermionized exactly at $K_c=K_s=1/2$. We therefore find
explicitly topological superconductivity in the charge sector, described by the Kitaev model of $p$-wave pairing in spinless fermion systems,
and bond-ordered insulator in the spin sector, described by the spinless SSH model. Such topological properties are expected to hold in a
considerable regime of $J_z/J$, but an exact phase diagram is difficult to draw, given the approximations made during bosonization of ${\cal
H}_{t-Jz}$ and ${\cal H}_p$ (specifically when dealing with the no-double occupation constraint).


\begin{thebibliography}{26}
\bibitem[{not({\natexlab{a}})}]{note1}
 \bibinfo{note}{When a positive Hubbard $U$ term ($U>2W$)
  is introduced, the degeneracy may be reduced to 2-fold, resulting solely from
  topological `superconductivity' in the spin sector, as the charge sector
  described by Mott insulator is topologically trivial (a paper in
  preparation.)}.
\bibitem[{not({\natexlab{b}})}]{note2}
\bibinfo{note}{For $t=|M|$, we find $\gamma_L=2c_{1\ua}^\dagger c_{1\da}+{\rm h.c}$
  and $\gamma_R=2i c_{N\ua}^\dagger c_{N\da}P_S+{\rm h.c}$, with $P_S=\prod_{j=1}^{N}e^{i\pi s_{j}^z}$ the spin parity.
  They satisfy $\{\gamma_L, \gamma_R\}=0$ and $\gamma_{E=L/R}^2=P_G$, where
  $P_G=\sum_{E=L/R,\sigma=\ua/\da}\ket{E_\sigma}\bra{E_\sigma}$ is the projection operator for the ground state manifold. }
\end{thebibliography}

\end{document}